\documentclass[%
 reprint,
showpacs,
 amsmath,amssymb, 
 aps,
prl,
]{revtex4-1}

\usepackage{graphicx}
\usepackage{dcolumn}
\usepackage{bm}

\usepackage{amsmath}
\usepackage{amsfonts}
\usepackage{amssymb}
\usepackage{epsfig}
\usepackage[pdftex]{color}
\usepackage[latin1]{inputenc}
\usepackage[T1]{fontenc}
\usepackage{lmodern}
\usepackage{bbm,array,theorem,enumerate,stmaryrd}
\usepackage{fancyhdr}
\usepackage{amsmath,amsxtra,amscd,amssymb,latexsym,comment,lineno,rotating}
\usepackage{textcomp}
\usepackage[toc,page]{appendix}
\usepackage{etoolbox}
\usepackage{setspace}
\usepackage{subfigure}
\usepackage{stfloats}
\usepackage{placeins}

\begin{document}

\preprint{APS/123-QED}

\title{Controlling the potential landscape and normal modes\\of ion Coulomb crystals by a standing wave optical potential}

\author{Thomas Laupr\^etre$^1$, Rasmus B. Linnet$^1$, Ian D. Leroux$^{1,2}$, Haggai Landa$^{3}$, Aur\'elien Dantan$^1$}

\author{Michael Drewsen$^1$}%
 \email{drewsen@phys.au.dk}
\affiliation{%
 $^1$Department of Physics and Astronomy, Aarhus University, DK-8000 Aarhus C, Denmark \\
$^2$National Research Council Canada, Ottawa, Ontario, Canada K1A 0R6 \\
$^3$Institut de Physique Th\'{e}orique, Universit\'e Paris-Saclay, CEA, CNRS, 91191 Gif-sur-Yvette, France
}%

\date{\today}

\begin{abstract}
Light-induced control of ions within small Coulomb crystals is investigated. By intense intracavity optical standing wave fields, subwavelength localization of individual ions is achieved for one-, two-, and three-dimensional crystals. Based on these findings, we illustrate numerically how the application of such optical potentials can be used to tailor the normal mode spectra and patterns of multi-dimensional Coulomb crystals. The results represent, among others, important steps towards controlling the crystalline structure of Coulomb crystals, investigating heat transfer processes at the quantum limit and quantum simulations of many-body systems.
\end{abstract}


\maketitle

{\it Introduction}---Trapped ions, laser cooled into long-range ordered structures, so-called Coulomb crystals, are a prime example of strongly correlated matter systems which are of broad relevance for plasma, solid-state and atomic physics~\cite{Dubin1999,Drewsen2015} as well as astrophysics, and whose unique properties make it possible to experimentally investigate various fundamental classical and quantum many-body systems~\cite{Porras2004,Blatt2012,Schneider2012,Bermudez2012,Britton2012,Chen2013,Genway2014,Schowalter2016,Bohnet2016,Kiethe2017}. They also represent well-controlled systems for cavity quantum electrodynamics~\cite{Herskind2009Realization,Albert2011,Casabone2013,Casabone2015,Begley2016}, quantum simulation~\cite{Porras2004,Blatt2012,Schneider2012,Hempel2018} and cold chemistry experiments~\cite{Willitsch2008,Willitsch2014}.

The additional application of optical potentials to trapped ions enlarges the range of possible applications. For instance, the interplay between optical and Coulomb forces can be used to investigate friction at the nanoscale~\cite{Garcia2007,Benassi2011,Pruttivarasin2011,Cetina2013,Bylinskii2015,Fogarty2015,Gangloff2015,Fogarty2015nanofriction,Bylinskii2016,Kiethe2017,Kiethe2018}, 
the dynamics of ions in quantum potentials~\cite{Bushev2004,Cormick2012} or energy transport in coupled oscillators systems~\cite{Pruttivarasin2011,Ramm2014,Ruiz2014,Freitas2015,Abdelrahman2016}.
Optical forces on ions can potentially also be exploited for pure optical trapping~\cite{Schneider2010,Huber2014,Schmidt2018}, useful for investigation of ultracold interactions between ions and neutrals~\cite{Grier2009,Zipkes2010,Schmid2010,Cetina2012,Huber2014}.

So far, experimental investigations of ion dynamics in optical lattices have been limited to single ions~\cite{Katori1997,Enderlein2012,Linnet2012,Schmiegelow2016} or small one-dimensional crystals~\cite{Karpa2013,Bylinskii2015,Gangloff2015,Bylinskii2016,Begley2016} in radiofrequency traps. In this Rapid Communication we investigate the control of the potential landscape and normal modes of Coulomb-interacting particles in multi-dimensional Coulomb crystals by a standing wave optical potential. Depending on the relative strengths of the Coulomb and optical forces various regimes of interest may be considered: {\it (i)} When the optical potential is stronger than the Coulomb and trapping potentials, the ions can be pinned along the standing wave field direction. We investigate this regime experimentally by demonstrating simultaneous subwavelength localization of up to eight $^{40}$Ca$^+$ ions in one-, two- and three-dimensional Coulomb crystals. In addition to the abovementioned applications, the results are promising, {\it e.g.}, for inhibiting spontaneous, thermally-activated crystalline structure changes observed in large Coulomb crystals~\cite{Drewsen2012}, as optical lattice potentials such as those applied here have been theoretically predicted to stabilize and enable deterministic control of their crystalline structure~\cite{Horak2012}. {\it (ii)} When the optical induced forces are comparable to those of the trapping potential, the application of the optical potential can be used to tailor, in a static or dynamic fashion, the normal modes of the crystals. We investigate this regime numerically and show the occurence of non-trivial normal mode dynamics for the multidimensional structures experimentally realized. Such a tailoring opens a new playground for fundamental investigations of energy transfer processes at the atomic scale, in which dimensionality plays a role~\cite{Ruiz2014,Freitas2015}.  It could also have important applications in connection with quantum many-body simulations with ion crystals in Penning traps \cite{Wang2013,Richerme2016,Stutter2018} or Paul traps \cite{Nath2015,Yoshimura2015,Ding2017parametric,Ding2017refrigerator}, for which using standing wave optical potentials for tailoring the normal modes of two-dimensional crystals~\cite{Mitchell1998,Mavadia2013,Stutter2018} would be a complementary alternative to the application of travelling waves~\cite{Britton2012,Bohnet2016} and potentially enlarge the toolbox for engineering effective spin-spin interactions.

\begin{figure}[h]%
\includegraphics[width=\columnwidth]{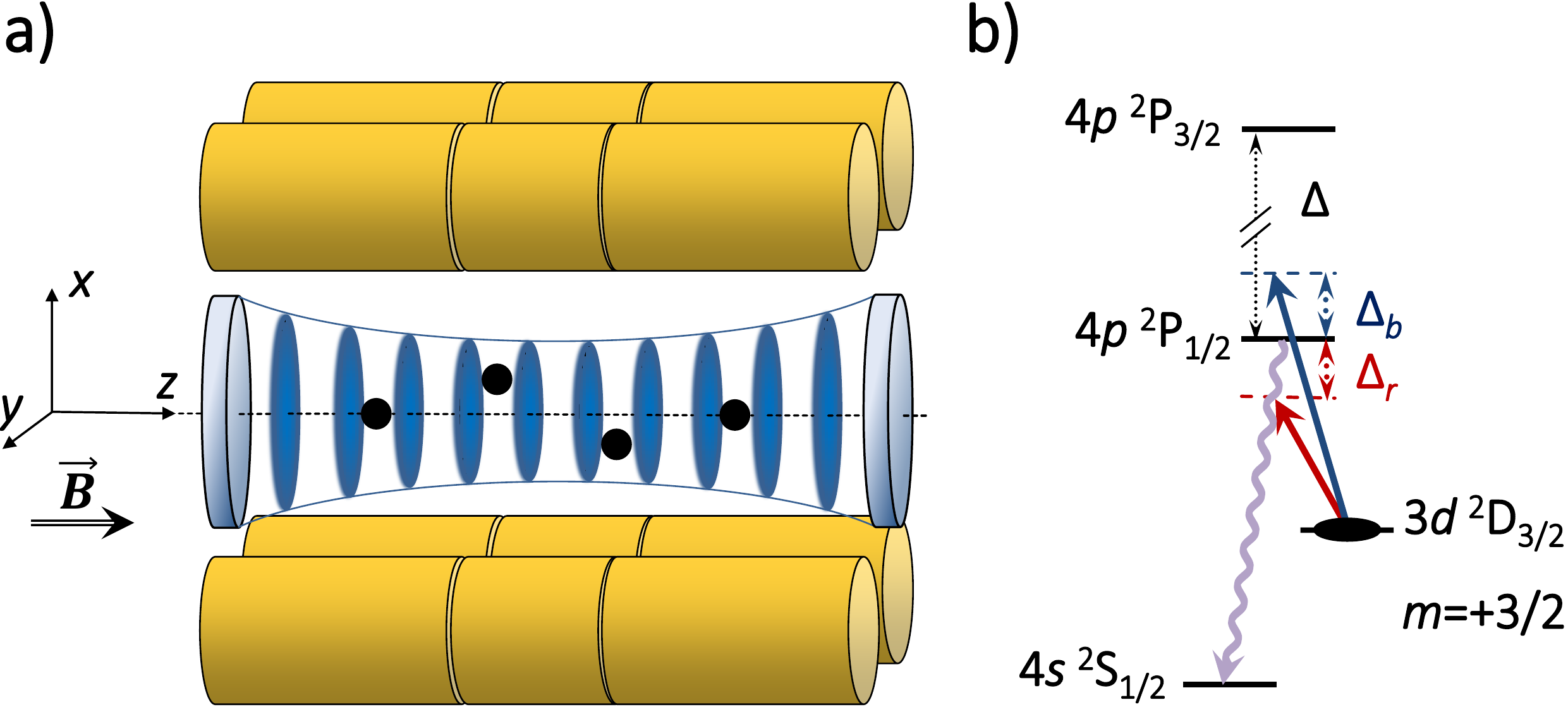}%
\caption{(color online).
(a) Experimental setup: ions trapped in a linear Paul trap and arranged in a Coulomb crystal are pinned along the $z$-direction by an intracavity standing wave field.
(b) Relevant energy levels in $^{40}$Ca$^+$. The long blue and short red straight arrows represent excitations by blue- or red-detuned standing wave fields. The purple wavy arrow represents the subsequent emitted and detected photon. $\Delta/(2\pi)\sim6.7$~THz is the frequency difference between the P$_{1/2}$ and P$_{3/2}$ states.}
\label{fig1}%
\end{figure}

{\it Experimental setup}---In the experiments, a number of $^{40}$Ca$^+$ ions is produced and confined in a linear Paul trap as described in detail in~\cite{Herskind2008,Herskind2009Positionning}  (Fig.~\ref{fig1}).
The trap operates at a radiofrequency (rf) of $\sim3.98$ MHz, with axial and radial trap frequencies in the range 70-110 kHz and 180-400 kHz, respectively.
The ions are first Doppler-cooled for $62\;\mu\mathrm{s}$, then optically pumped for $75\;\mu\mathrm{s}$ to the $|\mathrm{D}_{3/2},m=+3/2\rangle$ state (>98\% efficiency per ion on average) by the combined application of light fields close to resonance with the S$_{1/2}\rightarrow$ P$_{1/2}$ (397 nm) and D$_{3/2}\rightarrow$ P$_{1/2}$ (866 nm) transitions in the presence of a 2.2 G bias magnetic field along the $z$-axis (see~\cite{Herskind2009Realization}).
The resulting Coulomb crystal has a typical inter-ion distance of the order of $\sim20$ $\mu$m.
A 11.8 mm-long linear Fabry-Perot cavity with moderate finesse ($\sim3000$) and waist radius $\sim37$ $\mu$m allows for the generation of a standing wave along the $z$-axis with intensity up to $\sim500$ kW/cm$^2$ at the center of the trap.
Ions are positioned at the absolute center of the optical cavity following the method of~\cite{Linnet2014}.
After switching off the optical pumping fields, a $\sigma^-$-circularly polarized standing wave field detuned either to the blue or the red side of the D$_{3/2}\rightarrow$ P$_{1/2}$ transition by $\pm 0.76\;\mathrm{THz}$ is ramped up for $2\;\mu\mathrm{s}$ ~\cite{Linnet2012} and held at its maximum level for $1\;\mu\mathrm{s}$.
An independent and absolute calibration of the lattice potential depth experienced by a single ion as a function of the intensity transmitted out of the cavity is used as a reference~ \cite{Laupretre2019}, and a maximum lattice depth $T_{\textrm{latt}}\sim25$ mK, corresponding to a lattice vibrational frequency $\nu_{\textrm{latt}}\sim3.7$ MHz~\cite{frequency}, can be reached at this detuning in the limit of available laser power.
When an ion is excited to the P$_{1/2}$ state by the intracavity field, it leaves the $|\mathrm{D}_{3/2},m=+3/2\rangle$ state with 97\% probability by subsequently decaying to either the $|\mathrm{D}_{3/2},m=\pm1/2\rangle$ states  (3\%) or predominantly to the S$_{1/2}$ state (94\%) where it no longer interacts with the standing wave~\cite{m12}. The 397 nm photons scattered in the latter case are detected by an intensified CCD camera, gated to be active only during the $3\;\mu\mathrm{s}$ when the standing wave is applied.

\begin{figure}
\includegraphics[width=\columnwidth]{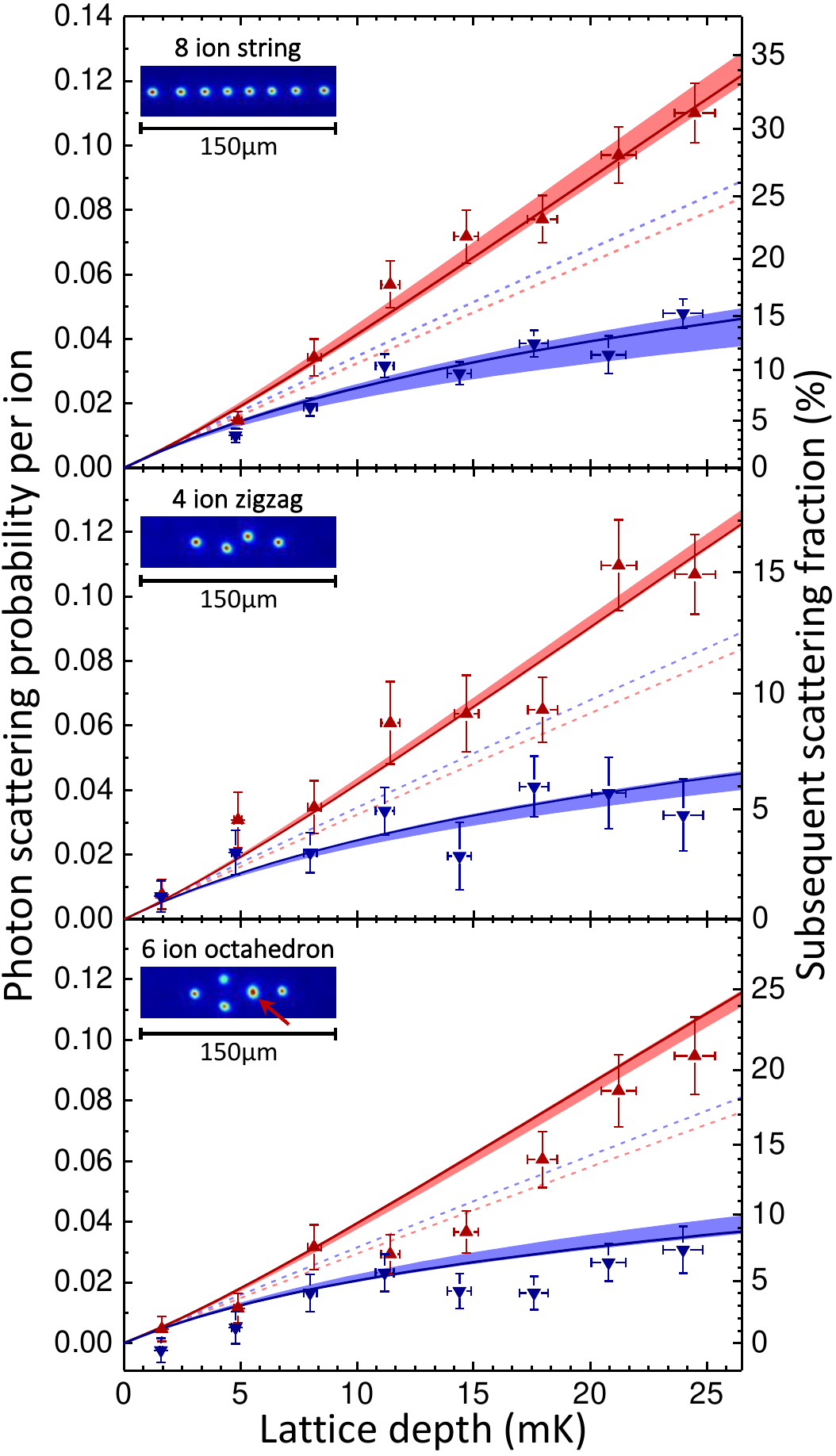}
\caption{(color online).
Photon scattering probability per ion and per sequence as a function of the optical lattice depth for an 8-ion string (top), a two-dimensional 4-ion zigzag crystal (center) and a three-dimensional 6-ion octahedron crystal (bottom).
The red up-triangles/blue down-triangles are experimental data points for the red-/blue-detuned lattices. Each data point corresponds to the repetition of approximately $2\times 10^6$ sequences.
The upper red and lower blue shaded areas are the theoretical scattering probabilities computed from the measured initial temperatures, with the thickness representing their error bar.
The upper red and lower blue continuous lines are the best results from single free parameter fits to the data.
The upper red and lower blue dashed lines show the theoretical scattering probabilities expected for delocalized ions; the slight asymmetry between scattering probabilities from red- and blue-detuned lattices is due to the non-negligible excitation to the P$_{3/2}$ state which is taken into account in the model.
The red arrow head on the fluorescence picture at the bottom indicates two out-of-plane ions overlapping.
}
\label{fig2}
\end{figure}

{\it Experimental results}---The measured scattering probability per ion and per experimental sequence is plotted in Fig.~\ref{fig2} as a function of the optical lattice depth for three different spatial configurations of the ions: a one-dimensional 8-ion string, a two-dimensional 4-ion zigzag crystal and a three-dimensional 6-ion octahedron crystal~\cite{octahedron}, obtained for axial and radial trap frequencies of (70,350), (85,170) and (105,190) kHz, respectively. The insets in Fig.~\ref{fig2} show fluorescence images of these crystals. To break the symmetry of the radial trapping potentials and improve the long term orientational stability of the zigzag and octahedron crystals, a small bias voltage resulting in a $\sim3\%$ difference in the radial frequencies was applied. In the experiments, the orientational or configurational changes of the crystals occurred at a rate of $\sim2$ s$^{-1}$ and $\sim4$ s$^{-1}$ for the zigzag and octahedron crystals, respectively.
As detailed in the Supplemental Material~\cite{suppl}, the initial average axial temperature of the ions in each configuration is evaluated by measuring their position variance based on the detected fluorescence images prior to the application of the optical lattice, and calculating numerically the frequency of the normal modes of motion from the axial and radial frequencies of the trap~\cite{Knunz2012,Rajagopal2016}. For the three crystals shown in Fig.~\ref{fig2}, the axial temperatures are found to be $3.6\pm 1.1$, $3.5\pm 0.5$, and $3.1\pm 0.5$ mK, respectively.

{\it Analysis and discussion}---Under these conditions, the initial position distribution of each ion extends over several lattice periods, so that the variation of the background trapping potential over one lattice period is small with respect to the initial thermal energy. Moreover, given the relatively large inter-ion distances used in this work, the lattice-induced forces quickly overcome the trapping and Coulomb forces along the axis as the lattice intensity is ramped up over a timescale comparable or shorter than the inverse of the secular axial frequencies. As such, the ions can be expected to independently localize close to the minima (maxima) of intensity of the blue (red)-detuned standing wave field, as demonstrated with a single ion in~\cite{Linnet2012}. In principle, the scattering dynamics of a multi-ion crystal are more complex than for a single ion, since, as soon as one ion gets excited by the standing wave field, it will decay with almost unit probability to a state which is not affected by the lattice (see above). However, in the limit where the probability of exciting more than one ion in each sequence is low, the detected scattering is essentially that of the first excited ion, thus providing information on its average position distribution inside the lattice potential before any depinning event. We therefore base the analysis on the single-ion model of Ref.~\cite{Linnet2012}, which, given the initial temperature of the ions, yields a prediction of the probability per ion to have scattered a photon during the application of the standing wave by determining the position distribution in the lattice at each instant~\cite{suppl}.

The predictions of the model based on the independently measured initial temperatures are shown in Fig.~\ref{fig2} (shaded areas) for each configuration~\cite{intensity}, and are generally observed to be in very good agreement with the experimental data points. For the octahedron crystal one observes though a general tendency of the data points to lie below the theoretical curves. This fact is likely due to a globally reduced measured scattering probability as a result of more frequent configurational changes (4 s$^{-1}$) and heating of the entire structure followed by slow recrystallization ($\gg$ ms) occurring over the long acquisition period of 5 mins per data point.
As an alternative analysis, we also show in Fig.~\ref{fig2} the best fit results of the theoretical model (Eqs.(1,2) in~\cite{suppl}) to the data points with the initial temperature only left as a single free parameter (solid lines). Such fits yield initial temperatures of $4.0\pm 0.5$, $3.9\pm 0.9$ and $2.7\pm 0.8$ mK, respectively, in very good agreement with the temperatures independently determined from the fluorescence images. The resulting uncertainties on the scattering probability (not shown) are similar to the ones obtained via the first analysis based on the initial temperature measurements. Note that this analysis does not include reduced scattering due to heating of the crystal either.

In addition, we show on the right scale of the graphs of Fig.~\ref{fig2} the subsequent scattering fraction, {\it i.e.}, the fraction of the signal that is due to the detection of any other photon than the first emitted one~\cite{suppl}. The single ion model is observed to give accurate predictions not only for all blue-detuned lattice data points, for which this fraction is always less than 15\%, but also for the red-detuned lattice data points, for which this fraction may be substantially higher. This may seem surprising since, as soon as one ion in a crystal scatters a photon from the standing wave field excitation, it decays with almost unit probability to a state which is not affected by the lattice and therefore changes the total potential energy of the system. The change in potential energy of the ions still affected by the optical lattice could possibly lead to a change in the localization process as their configuration changes, which would then alter subsequent scattering events. The experiment should, however, be insensitive to such effects for two reasons.
First, the interaction time with the lattice (3~$\mu$s in total) is never much longer than the period of the highest-frequency normal mode (2.6~$\mu$s in the lowest case of the 8-ion string).
Consequently, the system has too little time to relax and redistribute thermal energy between ions after a scattering event. Second, since the initial position distribution of each ion extends over several lattice sites, crystal distorsions due to pinning or depinning alter the potential energy of the system by only a fraction of the initial thermal energy.

Using the model predictions based on the experimentally determined initial temperatures~\cite{suppl}, we infer the "bunching parameter" $B=\langle\sin^2(kz)\rangle$, which represents the averaged normalized potential distribution ($B$ is 0 for ions perfectly localized at the potential minima and 1/2 for completely delocalized ions). For each of the configurations of Fig.~\ref{fig2} pinned inside the deepest blue-detuned lattice field ($\sim25$ mK) the inferred values are $0.22\pm0.03$, $0.22\pm0.01$ and $0.22\pm0.02$, respectively, which clearly indicate that subwavelength localization in the optical potential is achieved.

{\it Micromotion}---A general concern when experimenting with ion Coulomb crystals in linear rf traps is the rf-induced micromotion. Even for experiments with a single ion, where the ion in principle can be trapped on an rf nodal line, the ion's dynamics can be compromised due to resonances between the rf drive and the ion's oscillation in the optical potential~\cite{Schneider2010,Enderlein2012,Linnet2012}. For multi-dimensional crystals, some ions will furthermore always be positioned away from the rf-field-free nodal line with a resulting micromotion radial kinetic energy which can easily exceed the thermal energy of the ion by orders of magnitude~\cite{Berkeland1998,Landa2012}. For instance, the off-axis ions of the zigzag and octahedron crystals of Fig.~\ref{fig2} are estimated to have micromotion radial kinetic energies corresponding to $\sim160$ mK and $\sim800$ mK, respectively. Due to the general coupling between axial and radial motional degrees of freedom, it is hence not immediately clear that axial optical pinning of ions in such crystals can be achieved by realistic experimental optical potential depths of only $\sim25$ mK.  A theoretical analysis confirmed by numerical simulations at 0~K shows, however, that the micromotion kinetic energy due to the Coulomb interaction along the rf field-free direction is typically several orders of magnitude smaller than the radial micromotion kinetic energies in a linear Paul trap, both with and witout the optical potential present ~\cite{Landa2012,rfmodes}. For instance, for the 4-ion zigzag crystal, the amplitude of the axial micromotion of the external ions in a 25~mK-deep lattice is $\sim$4$\times10^{-3}$ times lower than the off-axis ion radial micromotion amplitude, corresponding to 0.35~\% of the lattice period and an additional axial kinetic energy of $\sim3~\mu$K. Residual micromotion due to potential parasitic axial components of the rf-field could be more of a concern. An experimentally determined upper bound of its amplitude in absence of lattice for the 8-ion string leads to a maximal associated kinetic energy less than 0.4~mK for the outer ions. According to the numerical simulations, this could lead in presence of a 25~mK-deep lattice potential to an axial micromotion amplitude of up to $\sim$17~\% of the lattice period (corresponding kinetic energy $\sim$8~mK) for these ions. Such an excess micromotion would, though, only contribute to the observed values of $\langle\sin^2(kz)\rangle$ by a few percent.

\begin{figure*}[h]
\centering
\includegraphics[width=\textwidth]{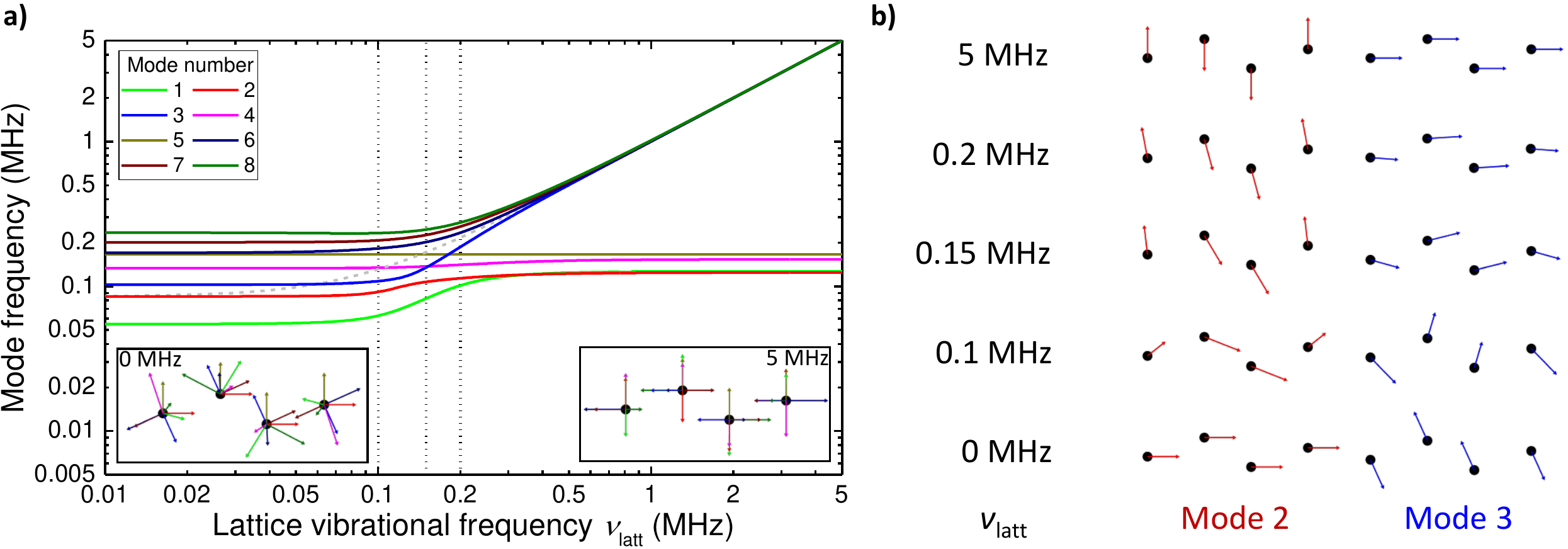}
\caption{(color online).
(a) Evolution of the frequency of the in-plane normal modes of the 4-ion zigzag configuration as a function of the lattice vibrational frequency $\nu_{\textrm{latt}}$. As a reference the center of mass mode frequency of a single ion is shown by the gray dashed line. The two insets show the modes coordinates (in relative units) without lattice (0 MHz) and in presence of the deepest lattice (5 MHz). (b) Evolution of the coordinates of the modes 2 and 3 {\color{blue} (whose frequencies in absence of lattice are respectively the second and third lowest, and shown by the red and blue lines in (a))}, for low and high lattice frequencies (0 and 5 MHz) and around the avoided crossing (0.1, 0.15 and 0.2 MHz, indicated by the vertical dotted lines in (a)).}
\label{fig3}
\end{figure*}

{\it Normal mode spectrum}---We finally turn to the prospect of tailoring the normal mode spectrum and patterns of Coulomb crystals by the application of an optical lattice potential. To this end, we have numerically calculated the parametric continuation of the normal modes as a function of the lattice depth for the three crystals presented in this paper. Results obtained for the 4-ion zigzag configuration are shown in Fig.~\ref{fig3}a, and for the other crystals in~\cite{suppl}. As observed, the mode spectrum changes dramatically when changing the lattice from being a weak perturbation to becoming the dominant axial confining potential. At the latter point, the axial and radial motional degrees of freedom nearly completely decouples, and all axial modes have identical frequencies. This degeneracy of the purely axial modes could, {\it e.g.}, facilitate the simultaneous resolved sideband laser cooling~\cite{Wineland1978} or cavity cooling~\cite{Fogarty2016optomechanical} of all axial modes to the ground state applying just a single-frequency laser. While the radial mode frequencies do not become degenerate for strong axial confinement, they cluster though within a significanly narrower distribution than without a lattice. Consequently, this scenario might facilitate ground state EIT cooling~\cite{Morigi2000} of all radial modes with fixed laser frequencies. Combining the two strategies for axial and radial cooling should be an interesting route to initialize small ion systems near their quantum mechanical ground state, which would be of relevance for investigating energy transport and thermodynamics in the quantum regime.

A perhaps even more remarkable feature is the complex evolution of the individual modes at moderate lattice depths. As an example, Figure~\ref{fig3}b illustrates how the center of mass mode (mode 2) and an essentially radial mode (mode 3) in absence of a lattice are switching nature as the lattice depth is adiabatically increased. One can envision several applications of this feature. For instance, both these two modes could potentially be cooled to the ground state by combining direct sideband cooling of only one of the two modes at a lattice depth outside the switching ("avoided crossing") area with fast non-adiabatically ramping of the lattice through the avoided crossing and adiabatically back again. Furthermore, by clever combinations of adiabatic and non-adiabatic rampings of the lattice potential, one might eventually be able to cool a larger number of modes. 

From a many-body physics perspective, the avoided crossings could be exploited to engineer hamiltonians with specific effective mode-couplings~\cite{Wang2013,Richerme2016,Nath2015,Yoshimura2015,Ding2017parametric,Ding2017refrigerator}, and one could, e.g., investigate the decoherence of superpositions of excitations of several of the modes converging to axial modes with the same frequencies at deep lattices after applying various ramping sequences. Since the highly non-trivial normal mode dynamics strongly depend on the dimensionality as well as the number of ions, as illustrated in~\cite{suppl}, the combination of ion Coulomb crystals with optical potential opens an exciting playground for investigating, exploiting and engineering complex interactions between motional/phonon modes.

{\it Conclusion}---We have experimentally demonstrated subwavelength localization of ions in multi-dimensional ion Coulomb crystals by applying intense optical standing wave fields. The fact that micromotion in these multidimensional ion crystal structures does not impede the lattice-induced localization is very promising not only for achieving deterministic control of the crystalline structure of cold charged plasmas~\cite{Horak2012}, but also for tailoring complex dynamics of their normal modes with applications within energy transport at the quantum limit as well as quantum many-body physics. As the specific coupling between modes achievable depends on dimensionality and number of ions, it expands the possibilities for tuning complex interactions in these strongly coupled systems, and may eventually be used in simulations of "artificial" molecules.

We acknowledge financial support from the European Commission (STREP PICC, FET Open TEQ and ITN CCQED projects), the Carlsberg Foundation, the Villum Foundation and the Sapere Aude Initiative from the Danish Council for Independent Research.


\bibliographystyle{apsrev}
\bibliography{Bibliography} 

\begin{thebibliography}{76}
\bibitem[{fre()}]{frequency}
\bibinfo{note}{The lattice frequency is defined as
  $\nu_{\textrm{latt}}=\frac{k}{2\pi}\sqrt{\frac{2k_\textrm{B}T_{\textrm{latt}}}{M}}$,
  where $k$ is the standing wave field wavenumber, $k_\textrm{B}$ the Boltzman
  constant and $M$ the mass of the ion.}
\bibitem[{m12()}]{m12}
\bibinfo{note}{The probability of interacting again with the lattice field, and
  subsequently heating the ion, after a decay to the
  $|\mathrm{D}_{3/2},m=+1/2\rangle$ state (2\%) is very low, due to the smaller
  Clebsch-Gordan coefficient and dipole element.}
\bibitem[{int()}]{intensity}
\bibinfo{note}{For the zigzag and the octahedron crystals, the lower light
  intensity experienced by the off-axis ions has also been accounted for.}
\end{thebibliography}

\end{document}


\preprint{APS/123-QED}

\title{Supplemental Material}

\author{Thomas Laupr\^etre$^{1}$}
	\email{thomas.laupretre@phys.au.dk}
\author{Rasmus B. Linnet$^1$, Ian D. Leroux$^2$, Haggai Landa$^{3}$,  Aur\'elien Dantan$^1$}
\author{Michael Drewsen$^1$}
	\email{drewsen@phys.au.dk}
\affiliation{%
 $^1$Department of Physics and Astronomy, Aarhus University, DK-8000 Aarhus C, Denmark \\
 $^2$National Research Council Canada, Ottawa, Ontario, Canada K1A 0R6 \\
$^3$Institut de Physique Th\'{e}orique, Universit\'e Paris-Saclay, CEA, CNRS, 91191 Gif-sur-Yvette, France}

\date{\today}

\pacs{37.10.Vz,37.10.Jk,03.67.Ac,37.10.Ty}

\maketitle

\tableofcontents

\section{Supplementary results}

\subsection{Experimental results}

\begin{figure*}[hbtp]
\centering
\includegraphics[width=\textwidth]{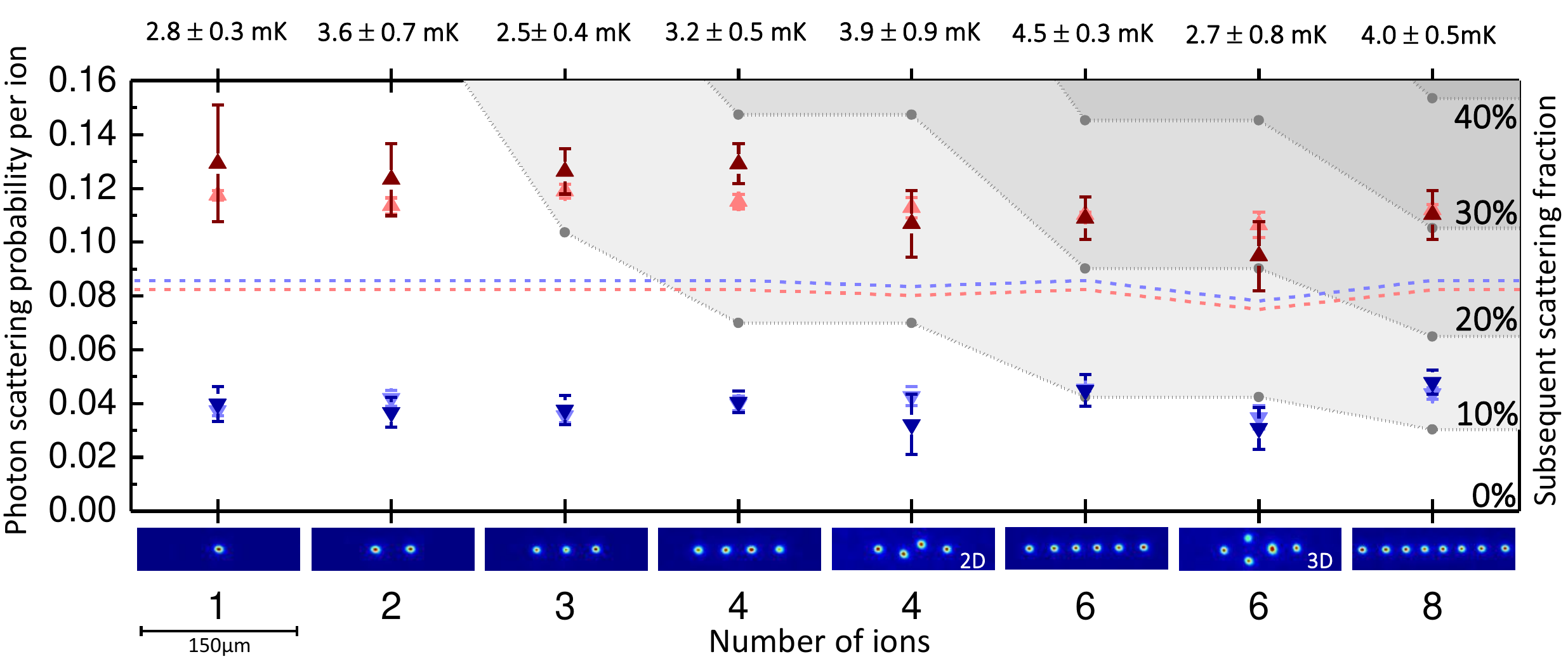}
\caption{(color online).
Photon scattering probability per ion for $\sim25\;\mathrm{mK}$-deep lattices as a function of the number of ions in the crystal. The red up-triangles/blue down-triangles are experimental data points for the red/blue detuned lattices.
The light-red and light-blue symbols show the corresponding results obtained from fitting to the theoretical model, while the red and blue dashed lines show the predicted theoretical results for delocalized ions.
The different shaded gray areas indicate the subsequent scattering fraction levels.}
\label{fig3}
\end{figure*}

To confirm the one-dimensional localization by the optical lattice potential, similar experiments were performed for crystals with various number of ions and structural configurations. Figure~\ref{fig3} shows the scattering probability per ion for both symmetrically blue- and red-detuned lattices with a fixed depth of $\sim~$25 mK as a function of the number of ions. The indicated initial temperatures are determined by fitting the average scattering probability from the standing wave field with the single ion theoretical model.

\subsection{Simulations}

\begin{figure}[hbtp]
\centering
\includegraphics[width=0.5\linewidth]{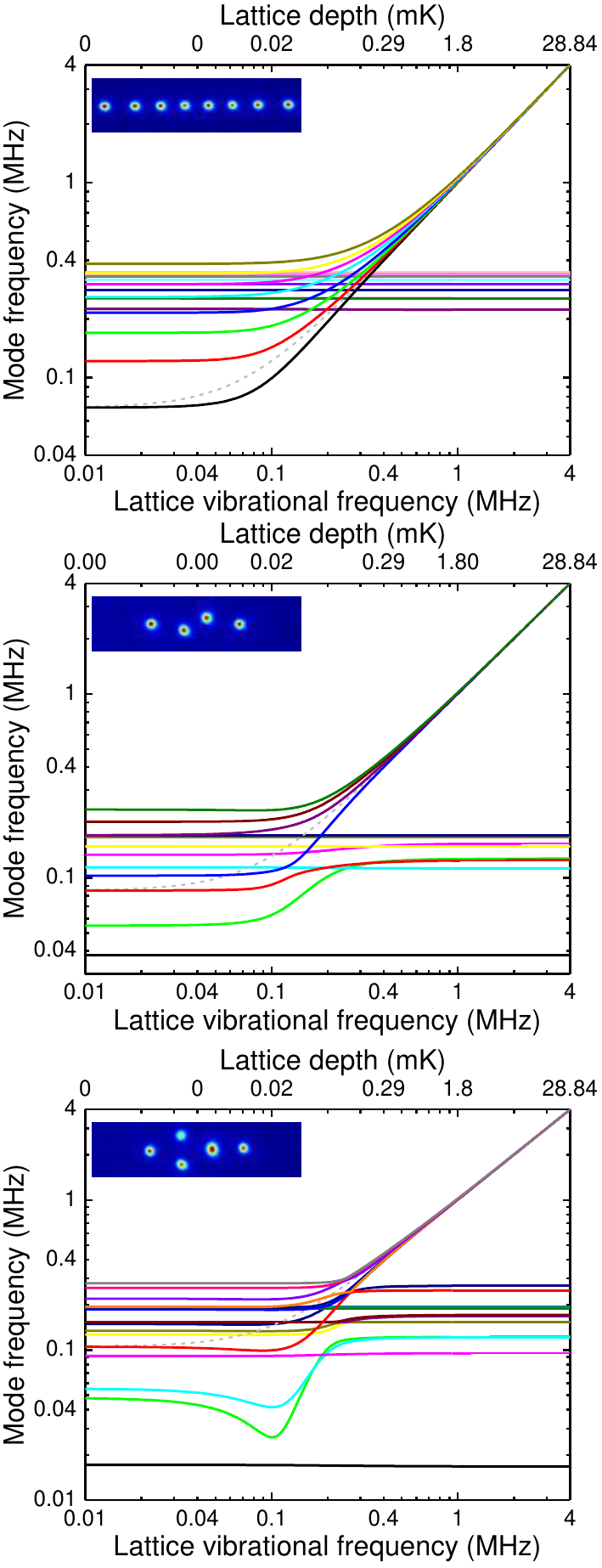}%
\caption{(color online).
Evolution of the frequency of the normal modes as a function of the lattice vibrational frequency at the bottom of the well for the three crystal configurations presented in the paper. The top scale gives the corresponding lattice depth.}
\label{Modes_All}
\end{figure}

Figure~\ref{Modes_All} shows,  for the three ion crystals experimentally realized (the 8-ion string, the 4-ion zigzag and the 6-ion octahedron), the evolution of the frequency of the normal modes as a function of the lattice vibrational frequency (depth). The evolution at low lattice depths is observed to strongly depend on dimensionality and number of ions,  in contrast to what happens at high lattice depths.

\newpage

\section{Optical lattice parameters}

The lattice potential along the longitudinal trap axis is $U(z)=U_0\sin^2(kz)$, with $k$ the optical field wavevector. The lattice depth, in units of temperature, is
\begin{equation*}
T_{\textrm{latt}}=\frac{\vert U_0\vert}{k_\textrm{B}},
\end{equation*}
where $k_\textrm{B}$ is the Boltzmann constant. The vibrational frequency (in Hz) of a particle of mass $M$ at the bottom of a well of the lattice is
\[
\nu_{\textrm{latt}}=\frac{k}{2\pi}\sqrt{\frac{2k_\textrm{B}T_{\textrm{latt}}}{M}}.
\]

\section{Analytical model of the scattering probability from the standing wave field}

\subsection{Single ion: adiabatically deformed pendulum \cite{Linnet2012,Linnet2014_PhD}}

Here is detailed the model used in \cite{Linnet2012} in the case of a single ion. To match the experimental conditions the model considers a single particle subject to a periodic potential in the coordinate, which is equivalent to a simple pendulum model. During adiabatic ramp-up of the lattice potential, the classical action $S$ of the ion's trajectory is conserved. The distribution of the action at any lattice depth $U_0$ is thus identical to the one of the initial ideal gas, which allows to calculate the energy distribution as well as the position distribution and the average scattering rate from the standing wave field at any time of the slow evolution of the lattice depth.

\
\paragraph{Action distribution}\

In the limit of a vanishing optical lattice potential depth as compared with the particle energy, the momentum $\pm p$ of the particle is independent of its position $z$ between the boundaries $\pm z_\mathrm{max}=\pm\pi/(2k)$, so the trajectory is a simple rectangle in phase space and the action of the system $\left. S\right\vert_{U_0=0}$ is directly proportional to the momentum
\[
\left. S\right\vert_{U_0=0}=\oint p\mathrm{d}z=\frac{2\pi}{k}\vert p\vert.
\]
The distribution of the action in the initial thermal gas at temperature $T_0$ is thus simply given by
\[
P(S)=\frac{k}{2\pi}P(\vert p\vert)=\frac{k}{2\pi}\frac{2}{\sqrt{2\pi Mk_\textrm{B}T_0}}e^{-\frac{p^2}{2Mk_\textrm{B}T_0}}=\sqrt{\frac{k^2}{2\pi^3Mk_\textrm{B}T_0}}e^{-\frac{k^2S^2}{8\pi^2Mk_\textrm{B}T_0}},
\]
and has the same form for any lattice depth $U_0$ when raised adiabatically. One can thus write
\[
P(S)=\frac{\nu_{\textrm{latt}}}{U_0}\frac{e^{-\frac{s^2}{4k_\textrm{B}T_0/U_0}}}{\sqrt{\pi k_\textrm{B}T_0/U_0}},
\]
where the dimensionless action $s=\frac{\nu_{\textrm{latt}}}{U_0}S$ has been introduced.

\
\paragraph{Energy distribution}\

Due to the symmetry of position $z$ and momentum $p$ the action $S(E)$ of a trajectory at a given energy $E$ is
\begin{align*}
S(E)=\oint p\mathrm{d}z=4\times\frac{\sqrt{2MU_0}}{k}\int_0^{kz_\mathrm{max}}\sqrt{\frac{E}{U_0}-\sin^2(kz)}\mathrm{d}(kz) =\frac{U_0}{\nu_{\textrm{latt}}}s(E).
\end{align*}
Here, $z_\mathrm{max}$ is the turning point of the particle's motion if its energy $E$ is lower than the lattice depth $U_0$, or the boundary $z_\mathrm{max}=\pi/2k$ if its energy $E$ is greater than the lattice depth $U_0$. The dimensionless action $s(E)$ has been introduced and can be written as a function of the complete elliptic integrals of the first kind $\mathcal{K}$ and second kind $\mathcal{E}$
\[
s(E)=\frac{4}{\pi}\times\left\{
\begin{array}{cc}
\mathcal{E}\left(\frac{E}{U_0}\right)-\mathcal{K}\left(\frac{E}{U_0}\right)\left(1-\frac{E}{U_0}\right) & E\leq U_0 \\
\sqrt{\frac{E}{U_0}}\mathcal{E}\left(\frac{U_0}{E}\right) & E>U_0. \\
\end{array}\right.
\]

Knowing the relation between action and energy, we deduce the energy distribution
\[
P(E)=P(S)\frac{\mathrm{d}S}{\mathrm{d}E}=\frac{e^{-s(E)^2\frac{U_0}{4k_\textrm{B}T_0}}}{U_0\sqrt{\pi k_\textrm{B}T_0/U_0}}\tau(E),
\]
where
\[
\tau(E)=\frac{\mathrm{d}s}{\mathrm{d}(E/U_0)}=\frac{2}{\pi}\times\left\{
\begin{array}{cc}
\mathcal{K}\left(\frac{E}{U_0}\right) & E\leq U_0 \\
\sqrt{\frac{U_0}{E}\mathcal{K}\left(\frac{U_0}{E}\right)} & E>U_0. \\
\end{array}\right.
\]

Note that the distribution has an integrable singularity in $U_0$.

\
\paragraph{Period of the trajectory}\

The period of a trajectory oscillating at a given energy $E<U_0$ in a lattice well can  be calculated as
\begin{align*}
T(E)=\oint \mathrm{d}t=\oint \mathrm{d}z/(\mathrm{d}z/\mathrm{d}t)=4\times\sqrt{\frac{M}{2k^2U_0}}\int_0^{kz_\mathrm{max}}\frac{\mathrm{d}(kz)}{\sqrt{E/U_0-\sin^2(kz)}}=\frac{\tau(E)}{\nu_{\textrm{latt}}},
\end{align*}
so that $\tau(E)$ represents the period of the trajectory normalized to that of small oscillations at the bottom of the lattice.

\
\paragraph{Position distribution}\

The position distribution of a trajectory at a given energy $E$ is:
\begin{align*}
P(kz\vert E) =\int_0^T\delta(kz-kz'(t))\frac{\mathrm{d}t}{T}=\frac{1}{\pi\tau(E)\sqrt{E/U_0-\sin^2(kz)}}.
\end{align*}
The total position distribution at a given energy depth $U_0$ and initial temperature $T_0$ can be deduced by numerical integration over the energy distribution
\[
P(kz)=\int_{U_0\sin^2(kz)}^{\infty}P(E)P(kz\vert E)\mathrm{d}E.
\]

\
\paragraph{Scattering rate}\

Solving the optical Bloch equations for a two-level system in steady state allows one to write the scattering rate on the P$_{1/2}\rightarrow\mathrm{S}_{1/2}$ transition from the lattice standing wave field excitation, $\Gamma_{sc}$, as a function of the position and the time-dependent Rabi frequency of the lattice field, $\Omega_{\mathrm{latt}}(t)=\sqrt{\frac{4\Delta_{\mathrm{latt}} U_0(t)}{\hbar}}$, when slowly varying the lattice depth $U_0(t)$, where $\Delta_{\mathrm{latt}}$ is the standing wave field detuning from the transition:
\[
\Gamma_{sc}(kz,\Omega_{\mathrm{latt}}(t))=\Gamma_{397}\Pi_e=\frac{\Gamma_{397}}{2}\frac{\frac{(\Omega_{\mathrm{latt}}(t)\sin(kz))^2}{2}}{\frac{\Gamma_{P_{1/2}}^2}{4}+\frac{(\Omega_{\mathrm{latt}}(t)\sin(kz))^2}{2}+\Delta_{\mathrm{latt}}^2}.
\]
$\Pi_e$ is the population in the P$_{1/2}$ state, $\Gamma_{P_{1/2}}$ is the total decay rate from the P$_{1/2}$ state and $\Gamma_{397}$ is the decay rate on the P$_{1/2}\rightarrow\mathrm{S}_{1/2}$ transition only.

The average over the position distribution slowly varying in time with the lattice depth $U_0(t)$ gives the scattering rate at any time $t$
\[
\langle\Gamma_{sc}\rangle(t)=\int_{-\pi}^{\pi}P(kz,t)\Gamma_{sc}(kz,\Omega_{\mathrm{latt}}(t))\mathrm{d}(kz).
\]

Because $\Gamma_{P_{1/2}}^2,\Omega_{\mathrm{latt}}^2<<\Delta_{\mathrm{latt}}^2$ for our experimental parameters, we actually use in order to speed up the numerical calculations
\begin{align}
\nonumber \langle\Gamma_{sc}\rangle(t) &=\Gamma_{397}\frac{\Omega_{\mathrm{latt}}(t)^2}{4\Delta_{\mathrm{latt}}^2}\int_{-\pi}^{\pi}P(kz,t)\sin(kz)^2\mathrm{d}(kz) \\
\nonumber &=\frac{I_{\mathrm{latt}}(t)}{\hbar\omega_{\mathrm{latt}}}\sigma_{397}\int_{0}^{1}P(\sin(kz)^2,t)\sin(kz)^2\mathrm{d}(\sin(kz)^2) \\
\nonumber &=\frac{I_{\mathrm{latt}}(t)}{\hbar\omega_{\mathrm{latt}}}\sigma_{397}\langle\sin(kz)^2\rangle(t) \\
&=\frac{I_{\mathrm{latt}}(t)}{\hbar\omega_{\mathrm{latt}}}\sigma_{397}\left\langle\frac{U}{U_0}\right\rangle(t).
\end{align}
We have introduced here the angular frequency of the lattice standing wave $\omega_{\mathrm{latt}}=2\pi\nu_{\mathrm{latt}}$, the antinode intensity of the lattice standing wave $I_{\mathrm{latt}}$, and the scattering cross section on the P$_{1/2}\rightarrow\mathrm{S}_{1/2}$ transition $\sigma_{397}$. It is indeed possible to write the average of the normalized potential at a given energy $E$
\begin{align*}
\langle \sin(kz)^2\rangle(E) &=\int_0^{min(\frac{E}{U_0},1)}P(\sin(kz)^2\vert E)\sin(kz)^2\mathrm{d}(\sin(kz)^2)\\
	&=\left\{
\begin{array}{cc}
1-\frac{\mathcal{E}\left(\frac{E}{U_0}\right)}{\mathcal{K}\left(\frac{E}{U_0}\right)} & E\leq U_0 \\
\frac{E}{U_0}\left(1-\frac{\mathcal{E}\left(\frac{U_0}{E}\right)}{\mathcal{K}\left(\frac{U_0}{E}\right)}\right) & E>U_0. \\
\end{array}\right.
\end{align*}
It can be numerically integrated over the energy distribution to give the averaged normalized potential for a given lattice depth $U_0$ and an initial temperature $T_0$
\[
B=\langle\sin^2(kz)\rangle=\left\langle\frac{U}{U_0}\right\rangle,
\]
also called \textit{bunching} parameter and used as a measure of the degree of localization.

\
\paragraph{Photon scattering probability}\

If we denote by $P(t)$ the probability to be in the initial state, we have
\[
\frac{\mathrm{d}P(t)}{\mathrm{d}t}=-\langle\Gamma_{sc}\rangle(t)P(t).
\]
Finally, the detected photon scattering probability after a time $t_0$ is the probability to have left the initial state after a time $t_0$, i.e the difference between the probability $P(0)$ to be in the initial state at the beginning and the probability $P(t_0)$ to be in the initial state at the time $t_0$:
\[
p(t_0)=P(0)-P(t_0)=P(0)\left(1-\exp\left[-\int_0^{t_0}\langle\Gamma_{sc}\rangle(t) dt\right]\right).
\]
Because of the optical pumping in the initial state during the preparation, we have $P(0)\simeq 1$ and:
\begin{equation}
p(t_0)=1-\exp\left[-\int_0^{t_0}\langle\Gamma_{sc}\rangle(t) dt\right].
\end{equation}

\subsection{Generalization to few ions crystals and potential deviations from the single ion model}

In the main text, we make use of the scattering probability per ion and per sequence. In case of small strings, the scattering probability is considered identical for each ion and the model is thus strictly identical to the single ion model. In case of the zigzag and octahedron crystals, the scattering of indivual ions is calculated with the single ion model in which the lattice depth is adjusted for each ion depending on its radial position in the beam profile, and the presented curves are the scattering probability averaged over the ions. 

As soon as one ion in a crystal scatters a photon from the standing wave field excitation, it decays with almost unit probability to a state which is not affected by the lattice. Thereby, it changes the total potential energy of the system.
The change in potential energy of the ions still affected by the optical lattice could possibly lead to a change in the localization process as their configuration changes, which would then alter subsequent scattering events.

This experiment should, however, be insensitive to such effects for two reasons.
First, the interaction time with the lattice (3~$\mu$s in total) is never much longer than the period of the highest-frequency normal mode (2.6~$\mu$s in the lowest case of the 8-ion string).
Consequently, the system has too little time to relax and redistribute thermal energy between ions after a scattering event.
Second, since the initial position distribution of each ion extends over several lattice sites, crystal distorsions due to pinning or depinning alter the potential energy of the system by only a fraction of the initial thermal energy.

Another possible complexity is imperfect optical pumping, which leads to an exponential decrease with $N$ of the probability to have initially all ions in the state addressed by the lattice.
However, this effect is less critical than that of a sudden depinning, since the lattice is slowly raised in presence of the unaffected ions, and it has observed to only reduce the scattering probability linearly.

\section{Subsequent scattering fraction}

We assume binomial statistics of the scattering of the ions in an $N$-ion crystal with an identical scattering probability per ion $p$. The probability for the ions in the crystal to scatter $N_p$ photon is given by 
\[
P_N(N_p)=\left(^N_{N_p}\right)p^{N_p}(1-p)^{N-N_p},
\]
so the average number of scattered photons is $N\times p$. The probability to scatter no photon at all is then $P_N(0)=(1-p)^{N}$. The complementary of $P_N(0)$ is the probability to scatter at least 1 photon: $P_N(N_p\geq1)=1-(1-p)^{N}$. As the first photon emitted exists as soon as some photon has been emitted, the average number of first emitted photons is $1\times P_N(N_p\geq1)=1-(1-p)^{N}$. The subsequent scattering fraction $f$ of signal due to the detection of any ohter photon than the first one is finally given by
\begin{equation*}
f=1-\frac{1-(1-p)^{N}}{Np}.
\end{equation*}

\section{Micromotion kinetic energy}

For any ion located at a distance $r_0$ from the rf-field free point in a ``non peculiar'' crystal, i.e. a crystal where the pseudo-potential limit correctly predicts the mean positions of the ions, the amplitude of excess micromotion $A_{\mu,u}$ along any direction $u$ is identical to the single ion case~\cite{Landa2012}, and is given as a function of the parameter $q_u$ of the trap along the considered direction by
\[
A_{\mu,u}=r_0\frac{q_u}{2}.
\]
The associated average kinetic energy for a particule of mass $M$ is then given by
\[
E_{\mu,u}^{\textrm{kin}}=\frac{1}{4}M\Omega_{rf}^2A_{\mu,u}^2\equiv \frac{1}{2}k_{\textrm{B}}T_{\mu,u}^{\textrm{kin}},
\]
where $T_{\mu,u}^{\textrm{kin}}$ represents the temperature associated to the driven micromotion energy.

\section{Determination of the initial temperature}

The initial temperature of the ions is estimated from fluorescence pictures taken at the end of the Doppler cooling part of the sequence. We follow the method developped in~\cite{Norton2011,Knunz2012} for single ions and for strings of ions in~\cite{Rajagopal2016} by introducing the crystal's normal modes of motion~\cite{James1998}. We have extended this method to two- and three-dimensional crystals, which we detail below before tackling the experimental procedure.

\subsection{Theory}

\subsubsection{Normal mode decomposition}

We consider an $N$-ion crystal in the pseudo-potential consisting of an axial harmonic trap with angular frequency $\omega_z$ and a radial harmonic trap with angular frequency $\omega_r$. The temperature is assumed to be low enough that the motional amplitude of each ion is small compared to the average distance between ions. A first order Taylor expansion of the total potential can be written as the sum of the potential of uncoupled harmonic oscillators, which define the $3N$ normal modes associated with the three-dimensional motion.

The excursions of the $m$-th ion from its equilibrium position $(x_m^0,y_m^0,z_m^0)$ are denoted by $(\delta x_m,\delta y_m,\delta z_m)$, so that, in a given direction $u$, $u_m(t)=u_m^0+\delta u_m(t)$. Let $(b_l^p)_{1\leq l\leq3N}$ and $\lambda_p$ be the $3N$ coordinates and the eigenvalue of the $p$-th eigenvector of the matrix describing the approximated potential in the Lagrangian normalized by $\omega_z^2$. The amplitude $Q_p$ of the $p$-th mode with frequency $\omega_p=\omega_z\sqrt{\lambda_p}$ can then be written as
\[
Q_p=\sum_{m=1}^Nb_m^p\delta x_m+\sum_{m=1}^Nb_{N+m}^p\delta y_m+\sum_{m=1}^Nb_{2N+m}^p\delta z_m,
\]
With these notations we have in each direction
\begin{equation*}
\delta x_m =\sum_{p=1}^{3N}b_m^pQ_p,\hspace{0.3cm}
\delta y_m =\sum_{p=1}^{3N}b_{N+m}^pQ_p,\hspace{0.3cm}
\delta z_m =\sum_{p=1}^{3N}b_{2N+m}^pQ_p.
\end{equation*}

Because these normal modes are uncoupled, one has:
\begin{equation*}
\langle \delta x_m^2\rangle =\sum_{p=1}^{3N}(b_m^p)^2\langle Q_p^2\rangle,\hspace{0.3cm}
\langle \delta y_m^2\rangle =\sum_{p=1}^{3N}(b_{N+m}^p)^2\langle Q_p^2\rangle,\hspace{0.3cm}
\langle \delta z_m^2\rangle =\sum_{p=1}^{3N}(b_{2N+m}^p)^2\langle Q_p^2\rangle.
\end{equation*}
and,
\begin{equation*}
\langle \dot{\delta x_m}^2\rangle =\sum_{p=1}^{3N}(b_m^p)^2\langle\dot{Q}_p^2\rangle,\hspace{0.3cm}
\langle \dot{\delta y_m}^2\rangle =\sum_{p=1}^{3N}(b_{N+m}^p)^2\langle\dot{Q}_p^2\rangle,\hspace{0.3cm}
\langle \dot{\delta z_m}^2\rangle =\sum_{p=1}^{3N}(b_{2N+m}^p)^2\langle\dot{Q}_p^2\rangle.
\end{equation*}
In the most general case,  let $T_{m,u}$ be the $m$-th ion's temperature in the direction $u$ ($u=x,y,z$) and $T_p$ the temperature associated with the $p$-th normal mode. The preceding relations imply that
\begin{equation*}
T_{m,x} =\sum_{p=1}^{3N}(b_m^p)^2T_p,\hspace{0.3cm}
T_{m,y} =\sum_{p=1}^{3N}(b_{N+m}^p)^2T_p,\hspace{0.3cm}
T_{m,z} =\sum_{p=1}^{3N}(b_{2N+m}^p)^2T_p.
\end{equation*}
In general, the normal modes do not necessarily have the same temperature. As each mode is a harmonic oscillator, we have though the relation
\[
k_{\textrm{B}}T_p=M\langle\dot{Q}_p^2\rangle=M\omega_p^2\langle Q_p^2\rangle,
\]
and the variances of the spatial excursions in the three directions can be expressed as a function of the normal mode temperatures as
\begin{equation*}
\langle \delta x_m^2\rangle =\frac{k_{\textrm{B}}}{M}\sum_{p=1}^{3N}(b_{m}^p)^2\frac{T_p}{\omega_p^2},\hspace{0.3cm}
\langle \delta y_m^2\rangle =\frac{k_{\textrm{B}}}{M}\sum_{p=1}^{3N}(b_{N+m}^p)^2\frac{T_p}{\omega_p^2},\hspace{0.3cm}
\langle \delta z_m^2\rangle =\frac{k_{\textrm{B}}}{M}\sum_{p=1}^{3N}(b_{2N+m}^p)^2\frac{T_p}{\omega_p^2}.
\end{equation*}

\subsubsection{Thermal equilibrium}

We now assume complete thermal equilibrium of the system, {\it i.e.} that all the normal modes have the same temperature $T$. Because $\sum_{p=1}^{3N}(b_i^p)^2=1$ $\forall i$, this implies that all ions also have the same temperature $T$, identical in each direction. The variance of the spatial excursions become
\begin{equation*}
\langle \delta x_m^2\rangle =\frac{k_{\textrm{B}}T}{M}\sum_{p=1}^{3N}\frac{(b_{m}^p)^2}{\omega_p^2},\hspace{0.3cm}
\langle \delta y_m^2\rangle =\frac{k_{\textrm{B}}T}{M}\sum_{p=1}^{3N}\frac{(b_{N+m}^p)^2}{\omega_p^2},\hspace{0.3cm}
\langle \delta z_m^2\rangle =\frac{k_{\textrm{B}}T}{M}\sum_{p=1}^{3N}\frac{(b_{2N+m}^p)^2}{\omega_p^2}.
\end{equation*}
Let us define constants associated with each ion and direction
\begin{equation*}
\gamma_{m,x}^2 =\sum_{p=1}^{3N}\frac{(b_{m}^p)^2}{\lambda_p},\hspace{0.3cm}
\gamma_{m,y}^2 =\sum_{p=1}^{3N}\frac{(b_{N+m}^p)^2}{\lambda_p},\hspace{0.3cm}
\gamma_{m,z}^2 =\sum_{p=1}^{3N}\frac{(b_{2N+m}^p)^2}{\lambda_p},
\end{equation*}
so that one can write in the usual harmonic oscillator form
\begin{equation*}
\langle \delta x_m^2\rangle =\frac{k_{\textrm{B}}T}{M\omega_z^2}\gamma_{m,x}^2,\hspace{0.3cm}
\langle \delta y_m^2\rangle =\frac{k_{\textrm{B}}T}{M\omega_z^2}\gamma_{m,y}^2,\hspace{0.3cm}
\langle \delta z_m^2\rangle =\frac{k_{\textrm{B}}T}{M\omega_z^2}\gamma_{m,z}^2.
\end{equation*}
It is clear that ions in a string at a temperature $T$ have a lower position excursion than the single ion at the same temperature because $\gamma_{m,z}<1$ $\forall m$. Note that all variances have been expressed as a function of the frequency of the center of mass mode frequency in the axial direction, because the potential was initially normalized to it.

\subsection{Experimental analysis}

\subsubsection{Evaluating the $\gamma$ parameters}

\noindent $\bullet$ To obtain the values of the $\gamma$ parameters for a given configuration, {\it i.e.} for a certain number of ions and fixed axial and radial frequencies, we make use of a Matlab program to numerically calculate from the pseudo-potential the mode coordinates $(b_l^p)_{1\leq l\leq3N}$ and their eigenvalues $\lambda_p$. Given the experimental uncertainty on the trap frequencies, a minimization of the difference of the calculated ion positions with the measured positions on fluorescence pictures is performed as a function of the trap frequencies to obtain more precise values.
\\

\noindent $\bullet$ In the radial direction, the modes coordinates have to be projected to account for the fact that the image plane is at $45^{\circ}$ from the radial trap axes. If $r_m$ is the $m$-th ion's coordinate in the radial direction in the image plane, one has
\begin{align*}
r_m &=\frac{x_m+y_m}{\sqrt{2}},\\
\delta r_m &=\sum_{p=1}^{3N}\frac{b_m^p+b_{N+m}^p}{\sqrt{2}}Q_p=\sum_{p=1}^{3N}b_{m,rad}^pQ_p,\\
\langle \delta r_m^2\rangle &=\frac{k_{\textrm{B}}T}{M\omega_z^2}\gamma_{m,rad}^2,
\end{align*}
where the parameter $\gamma_{m,rad}$ is defined as
\begin{equation*}
\gamma_{m,rad}^2 =\sum_{p=1}^{3N}\frac{(b_{m,rad}^p)^2}{\lambda_p}=\sum_{p=1}^{3N}\frac{(b_m^p+b_{N+m}^p)^2}{2\lambda_p}.
\end{equation*}

\subsubsection{Picture analysis}

\noindent $\bullet$ The thermal distribution of a harmonic oscillator in presence of a damping force and Brownian motion, as it is the case in the final stage of cooling, is almost gaussian~\cite{Blatt1986}. All modes have thus a gaussian distribution, which implies that the distribution of each excursion $\delta u_m$ is also gaussian.
\\

\noindent $\bullet$ The fluorescence spot observed on a picture is in each direction the convolution of the distribution of $\delta u_m$ with the Point Spread Function of the imaging system, which we assume to be gaussian with variance $\sigma_{\textrm{res}}^2$. The final recorded spot is thus gaussian with a variance in each direction given by
\[
\sigma_{m,u}^2=\langle \delta u_m^2\rangle+\sigma_{\textrm{res}}^2=\frac{k_{\textrm{B}}T}{M\omega_z^2}\gamma_{m,u}^2+\sigma_{\textrm{res}}^2.
\]
The resolution of our imaging system is $\sigma_{\textrm{res,ax}}^2=2.23\pm0.02~\mu$m in the axial direction and $\sigma_{\textrm{res,rad}}^2=2.09\pm0.02~\mu$m in the radial direction, and the pixellisation of our images with the experimental magnification corresponds to $0.92~\mu$m/pixel.
\\

\noindent $\bullet$ In the analyzed direction, the spot of the ion is integrated along the orthogonal direction and fitted by a gaussian function. The analysis is performed with Matlab with a Trust Region algorithm and the error bars are given by the 95\% confidence bounds.
\\

\noindent $\bullet$ In principle, the analysis should be carried out by processing all ions and directions at once with the equilibrium temperature $T$ as sole free parameter. In practice, we analyze all ions only in the $z$-direction for the following reasons:
\begin{itemize}
\item For strings, the lack of coupling, at leading (quadratic) order, between the radial and longitudinal degrees of freedom makes it possible to consider thermal equilibrium in the longitudinal direction separately.
\item For 2D and 3D configurations, we experimentally obtain stable configurations by introducing a small bias voltage on one pair of diagonal electrodes in order to introduce an asymmetry between the two radial frequencies. The exact value of this asymmetry plays an important role when numerically calculating the radial spatial extensions. When the asymmetry is low, a mode with very low frequency corresponding to a large radial extension of the off-axis ions is present as a vestige of the degeneracy.
This mode tends to disappear with increasing asymmetry, thus reducing the spatial extension of ions in the radial direction. This phenomenon translates into relatively large variations of the $\gamma$ values for off-axis ions in the radial direction. As an example, for the 6 ion octahedron crystal, the superimposed off-axis ions can have their radial $\gamma$ value modified by 12\% when the relative difference of frequency is changed by 10\%. In contrast, the $\gamma$ values in the axial direction have variations of the order of $\sim10^{-4}$. We thus analyze ions in the axial direction only, and evaluate that the error in the temperature measurement due to the uncertainty on the $\gamma$ values should not exceed $\sim 10^{-3}$.
\\
\end{itemize}

\noindent $\bullet$ Typically, we observe that the application of this method leads to relative error bars on the initial temperature estimation of the order of 20\%. We discuss below potential sources of systematic errors whose effects are found to be within this experimental uncertainty.

\subsection{Systematic errors}

\subsubsection{Doppler cooling}

The damping force due to Doppler cooling could lead to frequency shifts of the normal modes. A very conservative estimate based on the detunings and Rabi frequencies of the cooling lasers gives a systematic relative error lower than $6\cdot10^{-3}$ on the temperature. 

\subsubsection{Effect of micromotion}

In each direction, the secular motion of the $m$-th ion is written as a function of the normal mode amplitudes
\[
u_{m,\textrm{sec}}(t)=u_m^0+\sum_{p=1}^{3N}b_{m_u+m}^pQ_p(t)=u_m^0+\delta u_{m,\textrm{sec}}(t),
\] 
where $m_u=0,N,2N$ depending on the direction.

According to \cite{Landa2012}, in an ideal trap and for a non peculiar crystal, the amplitude of the micromotion depends on the position $u$ and the $q_u$ parameter along the considered direction as $A_{\mu,u}=u\frac{q_u}{2}$. If the highest frequency of the secular motion is small compared to the trap rf frequency, the total trajectory of the $m$-th ion can be written at first order in $q_u$:
\[
u_m(t)=\left(u_m^0+\sum_{p=1}^{3N}b_{m_u+m}^pQ_p(t)\right)\left(1+\frac{q_u}{2}\cos(\Omega_{rf}t)\right)=\delta u_{m}(t)+u_m^0\left(1+\frac{q_u}{2}\cos(\Omega_{rf}t)\right).
\]

\noindent $\bullet$ The position distribution to consider is then modified compared to the ideal case as it is given by the distribution of $\delta u_{m}$ which includes the ordinary micromotion and the position distribution of the excess micromotion $u_m^0\frac{q_u}{2}\cos(\Omega_{rf}t)$.
\begin{itemize}
\item First, the part without excess micromotion $\delta u_{m}$ has a gaussian distribution and has its variance given by $\langle \delta u_m^2\rangle=\langle \delta u_{m,\textrm{sec}}^2\rangle\left(1+\frac{q^2}{8}\right)$.

In the case of a string along the trap axis, an experimental upper bound of $q_{z}$ has been determined to be $\sim 5\cdot 10^{-4}$ in the conditions of these experiments, which leads to a negligible change of the position distribution and a relative error of $\sim3\cdot10^{-8}$ on the temperature measurement.

In the case of 2D or 3D crystals, since $q_z$ is very small and because of the coupling between degrees of freedom, the value to consider along the axial direction is the effective parameter $q'_z\sim\left(\frac{q_{\textrm{rad}}}{4}\right)^2$ \cite{Landa2012}, whose value, for $q_{\textrm{rad}}\sim 0.14$, is $\sim 1.2\cdot 10^{-3}$ in our experimental conditions. This leads also in this case to a negligible change of the position distribution and a relative error of $\sim2\cdot10^{-7}$ on the temperature measurement.

\item Therefore, the relative broadening of the position distribution in presence of micromotion is for all presented cases dominated by the amplitude of the excess micromotion. For all our experiments, the excess axial micromotion amplitude does not exceed 20~nm for the most external ions, whereas the width of the spatial distribution is at least $\sim 1$ $\mu$m. A very conservative estimate would give a corresponding relative error on the temperature measurement of $\sim4\cdot10^{-2}$.
\end{itemize}
Since the excess micromotion amplitude is negligible in comparison with the spatial thermal distribution, the position distribution stays gaussian and one still has the relation:
\[
\langle (u_m-u_m^0)^2\rangle\approx\langle \delta u_m^2\rangle\approx\langle \delta u_{m,\textrm{sec}}^2\rangle=\sum_{p=1}^{3N}(b_{m_u+m}^p)^2\langle Q_p^2\rangle.
\]
Therefore, one still has in the thermal equilibrium hypothesis
\[
\langle (u_m-u_m^0)^2\rangle\approx\frac{k_{\textrm{B}}T}{M\omega_z^2}\gamma_{m,u}^{'2},
\]
with
\[
\gamma_{m,u}^{'2}=\sum_{p=1}^{3N}\frac{(b_{m}^p)^2}{\lambda'_p},
\]
where $\gamma'_{m,u}$ now includes any shift of frequency of the normal modes due to micromotion in the eigenvalue $\lambda'_p=\omega_p^{'2}/\omega_z^2$.
\\

\noindent $\bullet$  In the case of a string along the trap axis, the frequencies of the axial normal modes are unchanged because there is no coupling, to leading (quadratic) order, between radial and axial directions. In the case of 2D or 3D crystals, the frequencies of the normal modes will in general be shifted, as compared to the pseudo-potential case, because of the coupling between degrees of freedom~\cite{Landa2012}.
The exact values of the modes frequencies could in principle have quite an influence on the temperature results, and only a complete Floquet-Lyapunov computation would tell about the relative error made on the frequencies. Because we have low $q_{\textrm{rad}}$ values for our multi-dimensional crystals, we can expect the shift to be small \cite{Kaufmann2012}. As an example, if we consider a relative frequency shift of the order of $3\cdot10^{-2}$, such as observed in \cite{Kaufmann2012}, this would lead to a relative error on the temperature lower than $6\cdot10^{-2}$.
\\

\section{Calculation of the normal modes in presence of the optical lattice}

Using a Matlab program identical to the one providing the $\gamma$ values useful to the measurements of the temperature of the ions without lattice, we numerically calculate the normal modes coordinates and frequencies (i.e. eigenvalues) in presence of a supplementary periodic potential corresponding to the lattice. Because the equilibrium positions of the ions change with the depth of the lattice potential, it is increased small step by small step for which the numerical minimization is done using the equilibrium positions of the previous step as the initial condition.

\bibliographystyle{apsrev}
\bibliography{3DLoc} 